\documentclass[twocolumn,showkeys,amsmath,amssymb,superscriptaddress]{revtex4}
\usepackage{graphicx}
\usepackage{dcolumn}
\usepackage{bm}
\begin{document}
\title{Three Dimensional Lattice-Boltzmann Model for Electrodynamics}
\author{M. Mendoza}
\email{mmendozaj@unal.edu.co}
\affiliation{
  Simulation of Physical Systems Group, Ceiba-Complejidad, Universidad Nacional de Colombia, Departamento de Fisica,\\
  Crr 30 \# 45-03, Ed. 404, Of. 348, Bogot\'a D.C., Colombia\\
}
\affiliation{ ETH Z\"urich, Computational Physics for Engineering
  Materials, Institute for Building Materials, Schafmattstrasse 6,
  HIF, CH-8093 Z\"urich (Switzerland) }
\author{J. D. Mu\~noz}
\email{jdmunozc@unal.edu.co}
\affiliation{
  Simulation of Physical Systems Group, Ceiba-Complejidad, Universidad Nacional de Colombia, Departamento de Fisica,\\
  Crr 30 \# 45-03, Ed. 404, Of. 348, Bogot\'a D.C., Colombia\\
}

\date{\today}
\begin{abstract}
  In this paper we introduce a novel 3D Lattice-Boltzmann model that
  recovers in the continuous limit the Maxwell equations in materials.
  In order to build conservation equations with antisymmetric tensors,
  like the Faraday law, the model assigns four auxiliary vectors to
  each velocity vector. These auxiliary vectors, when combined with
  the distribution functions, give the electromagnetic fields. The
  evolution is driven by the usual BGK collision rule, but with a
  different form for the equilibrium distribution functions. This LBGK
  model allows us to consider for both dielectrics and conductors with
  realistic parameters, and therefore it is adequate to simulate the
  most diverse electromagnetic problems, like the propagation of
  electromagnetic waves (both in dielectric media and in waveguides),
  the skin effect, the radiation pattern of a small dipole antenna and
  the natural frequencies of a resonant cavity, all with 2\% accuracy.
  Actually, it shows to be one order of magnitude faster than the
  original FDTD formulation by Yee to reach the same accuracy. It is,
  therefore, a valuable alternative to simulate electromagnetic fields
  and opens lattice Boltzmann for a broad spectrum of new applications
  in electrodynamics.
\end{abstract}



\maketitle
\section{Introduction}
The simulation of electromagnetic fields inside materials is a
fundamental tool in optics and electrodynamics, even more when
boundary conditions and geometries are so complex that an analytical
solution is out of question. Some examples are the design of antennas
and the study of electrical discharges across inhomogeneous media. For
this purpose, several numerical methods have been implemented. A
typical and well known example is the FDTD (Finite-difference
time-domain) method, which solves the time-dependent Maxwell equations
through a finite differences scheme proposed by Yee \cite{n100, yee2,
  yee3}. The method works pretty well in a broad range of
applications, but its stability strongly depends on the mesh size and
the time step. Several alternatives have been introduced, e.~g. the
Chebyshev \cite{chebmethod} and the unconditionally stable FDTD
\cite{uncond}, just to name a few.

In the last two decades, the arrival of Lattice Boltzmann methods has
been used as an alternative for the simulation of partial differential
equations. Originally motivated as discrete realizations of kinetic
models for fluids \cite{n5,n38}, they have also been developed for the
simulation of diffusion \cite{flekkoy}, waves \cite{n101, chop} and
even quantum mechanics \cite{s02a,s02b,s93} and relativistic
hydrodynamics \cite{newLB}. Electromagnetic fields, in contrast, have
been introduced meanwhile by models for plasmas, mostly in the form of
magnetic diffusion equations in resistive magnetohydrodynamics (MHD).

The first two-dimensional LB model reproducing the resistive MHD
equations was developed by S.~Succi et.~al.~\cite{n39}. Here, the
authors showed how a LB for the Navier-Stokes equations could be extended to
include two-dimensional magnetic fields and made systematic studies
in order to investigate the efficiency in terms of computational time.
Later on, other LB models have come as alternatives\cite{n6,n10,n11}.
S. Chen et al. \cite{n6} studied linear and nonlinear phenomena using
a LB model and showed that their method is competitive with
traditional solution methods. In another work by D.O. Martinez et al.
\cite{n10}, a new LB model for resistive magnetohydrodynamics was
proposed where the number of moving vectors could be reduced from
$37$ (used in previous works) to $13$, dramatically decreasing the amount of
computational memory requested.  One of the first models for
magnetohydrodynamics in 3D was developed by Osborn \cite{n11}, where he
used $19$ vectors on a cubic lattice for the fluid, plus $7$ vectors
for the magnetic field, which makes a total number of $26$ vectors per
cell.  Subsequently, Fogaccia, Benzi and Romanelli \cite{n12}
introduced a 3D LB model for turbulent plasmas incorporating the
electric potential in the electrostatic limit.  Among the many works
on LBM for MHD, the one proposed by Paul Dellar \cite{dellar} deserves
a special attention for the present work. It introduces the curl of
the electric field (i.e.  the divergence of an antisymmetric tensor)
indirectly, by using a vector-valued distribution for the magnetic
field which obeys a vector Boltzmann BGK equation \cite{bouchut}, and
is only coupled with the fluid distribution function via the
macroscopic variables evaluated at the lattice points.

In a recent work\cite{nmiller}, the authors introduced a 3D LB model
that reproduces the two-fluids theory for plasmas. The model employs
39 independent vectors (including both D3Q19 velocity vectors for the
fluids, D3Q13 auxiliary vectors for the electric field and D3Q7
auxiliary vectors for the magnetic one) and uses in general four
density distribution functions per velocity vector: one for the
electrons, one for the ions and two for the electromagnetic fields in
vacuum. The curls associated with the Faraday and Amp\'ere laws in
vacuum are explicitly constructed by means of the auxiliary vectors
mentioned above. The model reproduces the Hartman flow and allows to
reconstruct the phenomenon of magnetic reconnection in the magnetotail
with the true mass ratio between electrons and ions. However, the
problem of reproducing the Maxwell equations in media is not addressed
in that work. The solution requires for doubling the distribution
functions dealing with the electromagnetic fields, as we will show in
the present manuscript, plus additional forcing terms to account for
the sources. Once the terms related with the plasma fields (electrons
and ions) are removed, the set of velocity vectors can be simplified
to D3Q13, because we do not have viscous terms (the relaxation time is
$\tau$$=$$1/2$ and the procedure is still stable as a consequence of
the linear nature of Maxwell equations). The result is a complete
model that successfully reproduces the behavior of electromagnetic
fields inside dielectrics, magnets and conductors and gathers
information about the current density, electric charge, and
electromagnetic fields everywhere. The model is second-order accuracy
in time and performs well in a wide range of traditional benchmarks,
with errors below 1\% in the fields.

Section \ref{LBmodel} describes the model, with the evolution rules
and the equilibrium expressions for the 50 density functions, plus the
procedure to compute for the charge and current densities and the
electric and magnetic fields. This section also reviews how the
auxiliary vectors and the equilibrium distributions achieve the
construction of the curls in the Maxwell equations. The Chapman-Enskog
expansion showing how these rules recover the electrodynamic equations
with second-order accuracy is developed in Appendix
\ref{ChapmanEnskog}. In order to validate the model, we simulate in
section \ref{results} the reflection of an electromagnetic pulse on
the frontier between two dielectric media, the propagation of
electromagnetic waves along a microstrip waveguide, the skin effect
inside a conductor, the radiation pattern of an oscillating electrical
dipole (including a comparison between LB and Yee)
 and the resonant responses of a cubic cavity. The
main results and conclusions are summarized in section
\ref{conclusion}.

\section{3D Lattice-Boltzmann Model for Electrodynamics}
\label{LBmodel}
In a simple Lattice-Boltzmann model \cite{n5} the $D$-dimensional
space is divided into a regular grid of cells. Each cell has $Q$
vectors $\vec{v}_i$ linking with the neighboring cells, and each
vector has associated a distribution function $f_i$. This distribution
function evolves according to the Boltzmann equation,
\begin{equation}{\label{boltzeq}}
  f_i(\vec{x}+\vec{v}_i,t+1)-f_i(\vec{x},t)= \Omega_i(\vec{x},t) \quad ,
\end{equation}
where $\Omega_i(\vec{x},t)$ is a collision term, usually taken as a
time relaxation to some equilibrium function, $f_i^{\rm eq}$. This is
known as the Bhatnagar-Gross-Krook (BGK) operator \cite{n13},
\begin{equation}{\label{collop}}
  \Omega_i(\vec{x},t)=-\frac{1}{\tau}(f_i(\vec{x},t)-f_i^{\rm
  eq}(\vec{x},t)) \quad ,
\end{equation}
where $\tau$ is the relaxation time. The equilibrium function is
chosen in such a way, that (in the continuum limit) the model
simulates the actual physics of the system.
\begin{figure}
  \centering
  \includegraphics[scale=0.5]{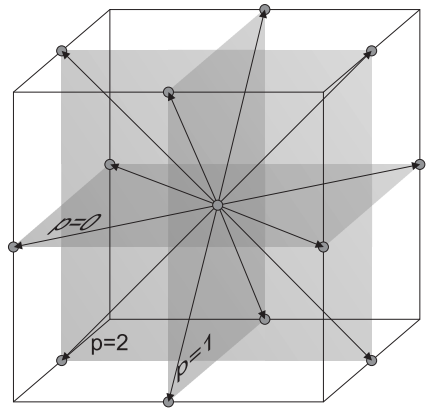}
  \caption{Cubic cell D3Q13 to model the Maxwell equations. The arrows
    represent the velocity vectors $\vec{v}_{i}^{p}$ and the electric
    field vectors $\vec{e}_{ij}^{p}$, where $p$ indicates the plane of
    location.}\label{d3q13}
\end{figure}
\begin{figure}
  \centering
  \includegraphics[scale=0.5]{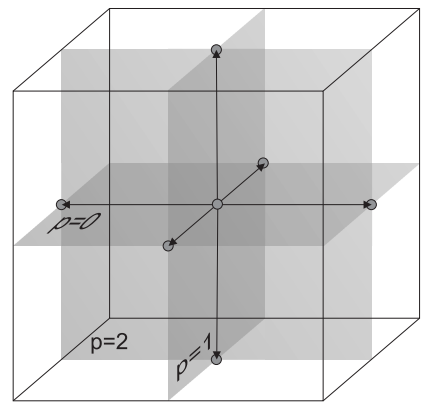}
  \caption{Cubic cell D3Q7 to simulate the magnetic field. The arrows
    indicate the magnetic vectors $\vec{b}_{ij}^{p}$.}\label{d3q7}
\end{figure}
\begin{figure}
  \centering
  \includegraphics[scale=0.5]{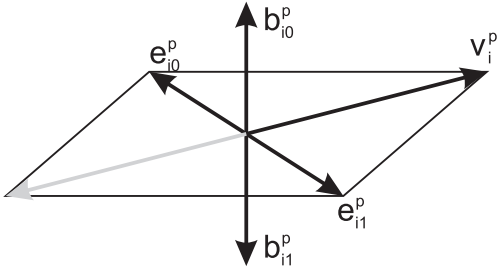}
  \caption{Index relationship between the velocity vectors and the
    electric and magnetic vectors.}\label{conexion}
\end{figure}

In our case, we need to reproduce the Maxwell equations in materials.
Two of them are the Faraday's law,
\begin{eqnarray}{\label{faradaylaw}}
  \frac{\partial \vec{B}}{\partial t}+\vec{\nabla} \times \vec{E}=0 \quad ,
\end{eqnarray}
and the Amp\`ere's law,
\begin{eqnarray}{\label{amperelaw}}
  \frac{\partial \vec{D}}{\partial t}-c^2\vec{\nabla} \times \vec{H}+\mu_0\vec{J}=0 \quad ,
\end{eqnarray}
where $\vec{B}$ is the induction field, $\vec{E}$ is the electric
field, $\vec{D}$ is the displacement field and $\vec{J}$ is the
current density. These are conservation equations including curls,
which can be written in tensorial form as
\begin{eqnarray}{\label{faradaylawtensor}}
  \frac{\partial \vec{B}}{\partial t}+\vec{\nabla} \cdot \Lambda =0
\end{eqnarray}
for the Faraday's law, and
\begin{eqnarray}{\label{amperelawtensor}}
  \frac{\partial \vec{D}}{\partial t}+\vec{\nabla} \cdot \Gamma
  +\mu_0\vec{J}=0
\end{eqnarray}
for the Amp\`ere's law, with the antisymmetric tensors
\begin{eqnarray}\label{tensorMax0}
  \Lambda=\left( \begin{array}{ccc}
    0 & -E_z & E_y\\
    E_z & 0 & -E_x \\
    -E_y & E_x & 0
  \end{array} \right) 
\end{eqnarray}  
and
\begin{eqnarray}\label{tensorMax1}
  \Gamma=\left( \begin{array}{ccc}
    0 & c^2B_z & -c^2B_y\\
    -c^2B_z & 0 & c^2B_x \\
    c^2B_y & -c^2B_x & 0
  \end{array} \right) 
  \quad . 
\end{eqnarray}  

Let us choose a cubic regular grid of lattice constant $\delta
x$$=$$\sqrt{2} c\delta t$, with $c$ the light speed in vacuum
($c$$\simeq$$3\times10^8 m/s$); that is, $c$$=$$1/\sqrt{2}$ in
normalized lattice units (time unit $=$$\delta t$, spatial unit
$=$$\delta x$), and the Courant-Fredericks-Levy criterion is automatically fulfilled.
There are $13$ velocity vectors per cell (figure
\ref{d3q13}), plus 13 different vectors for the electric field (figure
\ref{d3q13}) and 7 different vectors for the magnetic field (figure
\ref{d3q7}).  The velocity vectors are denoted by $\vec{v}_i^p$, where
$i$$=$$1,2,3,4$ indicates the direction and $p$$=$$0,1,2$ indexes the
plane of location (Fig. \ref{d3q13}). They have magnitude $\sqrt{2}$
(in lattice units) and lie on the diagonals of the planes.  In
components,
\begin{subequations}{\label{coord}}
  \begin{equation}
    \vec{v}_i^0=\sqrt{2}(\cos((2i-1)\pi/4),\sin((2i-1)\pi/4),0) \quad ,
  \end{equation}
  \begin{equation}
    \vec{v}_i^1=\sqrt{2}(\cos((2i-1)\pi/4),0,\sin((2i-1)\pi/4)) \quad ,
  \end{equation}  
  \begin{equation} 
    \vec{v}_i^2=\sqrt{2}(0,\cos((2i-1)\pi/4),\sin((2i-1)\pi/4)) \quad , 
  \end{equation}
\end{subequations}
They, plus the rest vector $\vec{v}_0$$=$$(0,0,0)$, give us $13$
vectors.  It is well known that the configuration D3Q13 has problems
to reproduce for fluids the local momentum conservation during
collision, due to a slackness in symmetry. Nevertheless, because we
are using $\tau$$=$$\frac{1}{2}$, there are not viscous terms like in
the fluid case, and D3Q13 can handle the symmetries we need for the
Maxwell equations, as shown below.

Associated to each velocity vector $\vec{v}_i^p$ there are two
electric $\vec{e}_{ij}^p$ and two magnetic auxiliary vectors
$\vec{b}_{ij}^p$ ($j$$=$$0,1$), as shown in Fig.\ref{conexion}.  They
are used to compute the electromagnetic fields. The electric vectors
are perpendicular to $\vec{v}_i^p$ and lie on the same plane $p$. The
magnetic vectors are also perpendicular to $\vec{v}_i^p$, but lying
perpendicular to the plane $p$. They are given by
\begin{eqnarray}\label{relationsfields}
  \begin{aligned}
    \vec{e}_{i0}^p=\frac{1}{2}\vec{v}_{[(i+2)\mod4]+1}^p & \quad ,
    \quad
    \vec{e}_{i1}^p=\frac{1}{2}\vec{v}_{[i\mod4]+1}^p \quad , \\
    \vec{b}_{ij}^p &= \vec{v}_i^p \times \vec{e}_{ij}^p \quad .
  \end{aligned}
\end{eqnarray}
The picture is completed by the null vectors $\vec{e}_{0}$$=$$(0,0,0)$
and $\vec{b}_{0}$$=$$(0,0,0)$.  With these definitions, there are $25$
electric field vectors, but only $13$ of them are different.
Similarly, there are $25$ magnetic field vectors, but only $7$ are
different. These vectors satisfy the following properties:
\begin{subequations}\label{propiedadesMax}
  \begin{equation}
    \sum_{i,p} \; v_{i\alpha}^p=0 \quad , 
  \end{equation}   
  \begin{equation}
    \sum_{i,j,p} \; e_{ij\alpha}^p=0 \quad , 
  \end{equation}   
  \begin{equation}
    \sum_{i,j,p} \; v_{i\alpha}^p e_{ij\beta}^p=0 \quad ,
  \end{equation}
  \begin{equation}
    \sum_{i,j,p} \; e_{ij\alpha}^p e_{ij\beta}^p=4\delta_{\alpha\beta} \quad ,
  \end{equation}
  \begin{equation}
    \sum_{i,j,p} \; v_{i\alpha}^p e_{ij\beta}^p e_{ij\gamma}^p=0 \quad ,
  \end{equation}    
  \begin{equation}
    \sum_{i,j,p} \; v_{i\alpha}^p e_{ij\beta}^p b_{ij\gamma}^p =
    4\epsilon_{\alpha\beta\gamma} \quad , \label{propiedadLeviCivita}
  \end{equation}    
  \begin{equation}
    \sum_{i,j,p} \; b_{ij\alpha}^p b_{ij\beta}^p = 8 \delta_{\alpha \beta}\quad ,
  \end{equation}    
\end{subequations}  
where $\epsilon_{\alpha\beta\gamma}$ are the components of the
Levi-Civita tensor, defined as usual: If two indexes are equal, the
component is zero, $\epsilon_{123}$$=$$1$ and any odd permutation of
indexes changes the component sign. The property
(\ref{propiedadLeviCivita}) is indeed the one that allow us to
construct conservation laws with antisymmetric tensors, as we will
show below.

The electromagnetic fields are computed from distribution functions
propagating from cell to cell with the velocity vectors $\vec{v}_i^p$.
There are four distribution functions associated with each non-zero
velocity vector, denoted by $f_{ij}^{p(r)}$ ($j$$=$$0,1$ and
$r$$=$$0,1$), plus two functions associated with the rest vector
$\vec{v}_0$, denoted by $f_0^{(r)}$. This makes $4\times12+2$$=$$50$
distribution functions.The macroscopic fields are computed as follows:
\begin{subequations}{\label{macros}}
  \begin{equation}
    \vec{D}=\sum_{i=1}^4 \sum_{p=0}^2 \sum_{j=0}^1 f_{ij}^{p(0)} \vec{e}_{ij}^p \quad ,
  \end{equation}
  \begin{equation}
    \vec{B}=\sum_{i=1}^4 \sum_{p=0}^2 \sum_{j=0}^1 f_{ij}^{p(1)} \vec{b}_{ij}^p \quad ,
  \end{equation}
  \begin{equation}
    \rho_c= f_{0}^{(0)} +\sum_{i=1}^4 \sum_{p=0}^2 \sum_{j=0}^1 f_{ij}^{p(0)}  \quad ,
  \end{equation}
  \begin{equation}
    \vec{E}= \frac{\vec{D}}{\epsilon_r} \quad ,  
  \end{equation}
  \begin{equation}
    \vec{H}= \frac{\vec{B}}{\mu_r} \quad ,  
  \end{equation}
  \begin{equation}
    \vec{J}= \sigma \vec{E} \quad ,  
  \end{equation}
\end{subequations}
where $\vec{D}$, $\vec{E}$ and $\vec{J}$ are subsidiary fields
representing the displacement field, the electric field and the total
current density before external forcing, respectively (the actual
mean fields, including external forcing, are described below). In
addition, $\vec{B}$ is the induction field, $\vec{H}$ is the magnetic
field, $\rho_c$ is the total charge density and $\epsilon_r$, $\mu_r$
and $\sigma$ are the relative dielectric constant, the relative
permeability constant and the conductivity for the medium,
respectively.  Notice that we define $\vec{D}$ and $\vec{H}$ just with
the relative constants $\epsilon_r$ and $\mu_r$, instead of the total
electromagnetic constants $\epsilon$$=$$\epsilon_r \epsilon_0$ and
$\mu$$=$$\mu_r \mu_0$.

The velocity vectors $\vec{v}_i^p$ are not mean to represent the
velocity of any particles (in contrast with the normal lattice
Boltzmann models for fluids), because in classical electrodynamics
there are not such particles, but they can be related with Poynting
vectors describing the momentum density flux of the electromagnetic
fields. This interpretation can be supported by looking at
Eqs.~\ref{macros} and Fig.~\ref{conexion}, and recognizing on the
figure the implicit cross product that defines the Poynting vector as
$\vec{S}$$=$$\frac{1}{\mu_r\epsilon_r}\vec{D}\times\vec{B}$\cite{n16}.

For the collision we adopt terms $\Omega_{ij}^{p(r)}$ and
$\Omega_{0}^{(r)}$ of the BGK form \cite{n13}
\begin{subequations}
  \begin{equation}
    \Omega_{ij}^{p(r)}=-\frac{1}{\tau}(f_{ij}^{p(r)}(\vec{x},t)-f_{ij}^{p(r)
      \rm eq}(\vec{x},t))\quad ,
  \end{equation}
  \begin{equation}
    \Omega_{0}^{(r)}=-\frac{1}{\tau}(f_{0}^{(r)}(\vec{x},t)-f_{0}^{(r)\rm eq
    }(\vec{x},t))\quad .
  \end{equation}
\end{subequations}
where the relaxation time is chosen $\tau$$=$$\frac{1}{2}$. Since
Maxwell equations are linear, this relaxation time does not produce
any instability.

In order to include the source term in the Ampere's law
(\ref{amperelaw}), we need of external forcing terms. These terms
are included by following the proposal of Zhaoli Guo, Chuguang Zheng
and Baochang Shi \cite{n30}, as follows:
\begin{eqnarray}{\label{lbe2}}
  \begin{aligned}
    f_{ij}^{p(r)}(\vec{x}+\vec{v}_i^p,t+1)-f_{ij}^{p(r)}(\vec{x},t)=&
    \Omega_{ij}^{p(r)}(\vec{x},t)+ T_{ij}^{p(r)} ,
  \end{aligned}
\end{eqnarray}
\begin{eqnarray}{\label{lbe4}}
  f_{0}^{(r)}(\vec{x},t+1)-f_{0}^{(r)}(\vec{x},t)= \Omega_{0}^{(r)}(\vec{x},t)
  + T_0^{(r)}\quad ,
\end{eqnarray}
where $T_{ij}^{(r)}$ and $T_0^{(r)}$ are forcing coefficients
($r$$=$$0,1$). These coefficients are defined by \cite{n30}
\begin{subequations}\label{coeffforce}
  \begin{eqnarray}
    \begin{aligned}
      T_{ij}^{p(r)}&= \biggl ( 1-\frac{1}{2\tau} \biggr)\biggl
      (\frac{1}{4}(\vec{e}_{ij}^p - \vec{E'})\cdot \vec{F} \\ &+
      \frac{3}{4}(\vec{e}_{ij}^p\cdot\vec{E'})(\vec{e}_{ij}^p\cdot \vec{F}) \biggr) \quad ,
    \end{aligned}
  \end{eqnarray}
  \begin{equation}
    T_0^{(r)}= \biggl ( 1-\frac{1}{2\tau} \biggr) \biggl
      (-\frac{1}{4}(\vec{E'})\cdot \vec{F} \biggr) \quad ,
  \end{equation} 
\end{subequations}
with $\vec{F}$ the external forcing. Since $\tau$$=$$1/2$,
$T_{ij}^{p(r)}$$=$$T_0^{(r)}$$=$$0$, and the forcing only appears in
the mean fields.  The mean electric field, $\vec{E'}$, and the mean
density current vector, $\vec{J'}$, are given by
\begin{eqnarray}
  \vec{E'}&=&\vec{E}-\frac{\mu_0}{4\epsilon_r}\vec{J'} \quad ,
      {\label{expandE}}\\ \vec{J'}&=&\sigma \vec{E'} \quad .
      {\label{currentequil}}
\end{eqnarray}
Replacing Eq.\eqref{expandE} into Eq.\eqref{currentequil} gives us the
mean density vector $\vec{J'}$ in terms of the subsidiary fields,
\begin{equation}{\label{expandEsub}}
  \vec{J'}=\frac{\sigma}{1+\frac{\mu_0 \sigma}{4\epsilon_r}}\vec{E} \quad .
\end{equation}

Finally, the equilibrium distribution functions for the
electromagnetic fields are given by
\begin{subequations}{\label{equilc}}
  \begin{equation}
    f_{ij}^{p(0) \rm eq}(\vec{x},t)=\frac{1}{16}\vec{v}_i^p \cdot
    \vec{J'}+\frac{\epsilon}{4}\vec{E'} \cdot
    \vec{e}_{ij}^{p}+\frac{1}{8\mu}\vec{B} \cdot \vec{b}_{ij}^{p} \quad ,
  \end{equation}
  \begin{equation}
    f_{ij}^{p(1) \rm eq}(\vec{x},t)=\frac{1}{16}\vec{v}_i^p \cdot
    \vec{J'}+\frac{1}{4}\vec{E'} \cdot \vec{e}_{ij}^{p}+\frac{1}{8}\vec{B}
    \cdot \vec{b}_{ij}^{p} \quad ,
  \end{equation}
  \begin{equation}
    f_{0}^{(0)\rm eq}(\vec{x},t)=f_{0}^{(1)\rm eq}(\vec{x},t)=\rho_c \quad ,
  \end{equation}
\end{subequations}
where the two sets ($r$$=$$0,1$) of equilibrium density functions
($f_{ij}^{p(r) \rm eq}$) are made explicit. This structure for the
equilibrium functions allows for the construction of conservative laws
with curls. Indeed, performing the Chapman-Enskog expansion of the BGK
evolution equation with these equilibrium functions, but multiplying
by $\vec{b}_{ij}^p$ \--- instead of the traditional multiplication by
the velocity vectors $\vec{v}_i^p$ (see Appendix \ref{ChapmanEnskog})
\--- before summing up over the index $i$, $j$ and $p$, we obtain the
time derivative of the magnetic field equals to the divergence of the
antisymmetric tensor (\ref{tensorMax1}). The tensor becomes
antisymmetric because this procedure builds up the Levi-Civita tensor,
as follows: The Chapman-Enskog expansion gives a velocity vector
$\vec{v}_i^p$, the equilibrium function contributes with a vector
$\vec{e}_{ij}^p$ and multiplying by $\vec{b}_{ij}^p$ and summing up
over $i$, $j$ and $p$ gives us $\epsilon_{\alpha\beta\gamma}
E'_\gamma$ for the tensor (Eq. \ref{propiedadLeviCivita}), that is the
Faraday's law.  Amp\`ere's law is obtained by multiplying by
$\vec{e}_{ij}^p$ (instead of $\vec{b}_{ij}^p$) and following the same
procedure.  This completes the definition of the lattice Boltzmann
model. The detailed proof that this LBGK model, via a Chapman-Enskog
expansion, recovers the Maxwell equations is shown in Appendix
\ref{ChapmanEnskog}.

The model reproduces the following equations:
\begin{subequations}\label{EMTeoricMaED}
  \begin{equation}
    \frac{\partial \rho_c}{\partial t}+\nabla \cdot \vec{J'}=0 \quad ,
  \end{equation}
  \begin{equation}
    \nabla \times \vec{E'}=-\frac{\partial \vec{B}}{\partial t} \quad ,
  \end{equation}
  \begin{equation}
    \nabla \times \vec{H}=\mu_0 \vec{J'}+\frac{1}{c^2}\frac{\partial
      \vec{D'}}{\partial t} \quad .
  \end{equation}
\end{subequations}
It is well known \cite{n16} that these three conservative laws implies the following expressions
for the divergence of the electromagnetic fields:
\begin{subequations}\label{EMTeoricMaEDder}
  \begin{equation}
     \frac{\partial}{\partial t} \left( \nabla \cdot
     \vec{D'}-\frac{\rho_c}{\epsilon_0} \right) =0 \quad ,
  \end{equation}
  \begin{equation}
   \frac{\partial}{\partial t} \left( \nabla \cdot \vec{B} \right) =0 \quad ;
  \end{equation}
\end{subequations}
that is, the other two Maxwell equations are reproduced if they are
satisfied by the initial condition (for more details see Appendix
\ref{ChapmanEnskog}). 

\section{Numerical Tests}
\label{results}
In order to validate our model, we have implemented several tests. Let
us start with two simple simulations showing that the model reproduces
the correct electromagnetic propagation with dielectrics and
conductors. We will construct more complex simulations afterward, like
the radiation pattern from an oscillating electric dipole, the wave
propagation on a waveguide, and the identification of the normal modes
in a cubic resonance cavity. Additionally, in this section, we compare
the results obtained by using LB and Yee methods for the case of the
radiation produced by an oscillating dipole.

\subsection{Dielectric Interface}
\begin{figure}
  \centering
  \includegraphics[angle=0, scale=0.33]{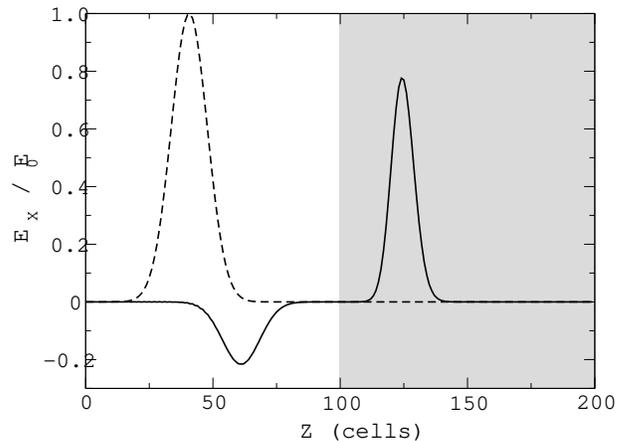}
  \caption{Electric pulse crossing a dielectric interface. The shadow
    zone is the dielectric medium, with dielectric constant
    $\epsilon_r$$=$$2.5$ and the other one corresponds to the vacuum
    ($\epsilon_r$$=$$1.0$). The curves are the intensity of the
    electric field at $t$$=$$0$ (dashed line), and at $t$$=$$140$
    (solid line). }\label{dielec0}
\end{figure}

As a first benchmark let us simulate the propagation of an
electromagnetic Gaussian pulse crossing a dielectric interface. For
this purpose, we took an uni-dimensional array of $L$ cells with
periodic boundary conditions in the $z$ coordinate and with each cell
being its own neighbor in both $x$ and $y$ directions. One half of the
simulation space, $z$$<$$L/2$, is vacuum ($\epsilon$$=$$\epsilon_0$)
and the other half, $z$$>$$L/2$, represents a dielectric medium with
relative dielectric constant $\epsilon_r$$=$$\epsilon /
\epsilon_0$$=$$2.5$. In order to avoid for abrupt changes on the
dielectric constant between two neighboring cells we choose the
following distribution of the permittivity:
 \begin{equation}
   \epsilon_{\rm r}=1.75 + 0.75\tanh(x-L/2) \quad .
 \end{equation}
 The functional form of the incident Gaussian electromagnetic pulse
 centered at $z_0$ is given by
 \begin{subequations}\label{pulsogauss}
   \begin{equation}
  \vec{E}=(E_0 \exp(-\alpha(z-z_0)^2),0,0) \quad ,
\end{equation}
\begin{equation}
  \vec{B}=(0,B_0 \exp(-\alpha(z-z_0)^2),0) \quad .
\end{equation}
\end{subequations}
The constant $\alpha$ fixes the pulse width, $E_0$ is the pulse
amplitude and the constant $B_0$ is related with $E_0$ according to
the relation $E_0$$=$$cB_0$, with $c$ the vacuum light speed.  For the
simulation we choose $L$$=$$200$, $c$$=$$1/\sqrt{2}$, $E_0$$=$$0.001$,
$\alpha$$=$$0.01$ y $z_0$$=$$40$ (in normalized units). The initial
condition and the electric field after $140$ time steps is shown in
figure \ref{dielec0}.

\begin{figure}
  \centering
  \includegraphics[angle=0, scale=0.58]{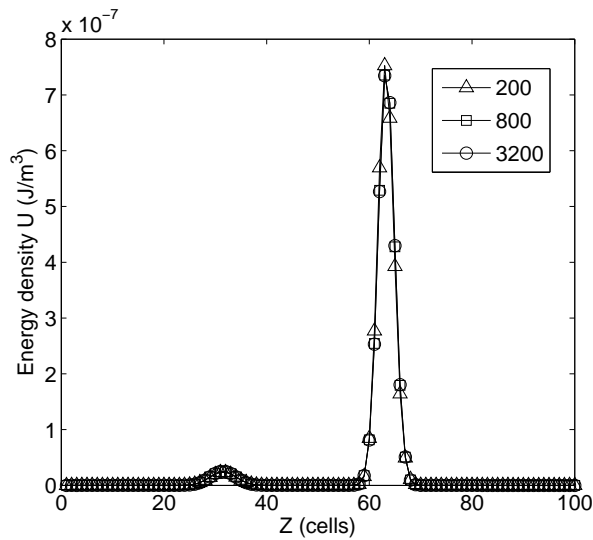}
  \caption{Energy density $U$ as a function $z$ at $t$$=$$140$ for an
    electric pulse crossing a dielectric interface for different grid
    resolutions: $200$, $800$, and $3200$ cells.}\label{energydensity}
\end{figure}

\begin{figure}
  \centering
  \includegraphics[angle=0, scale=0.58]{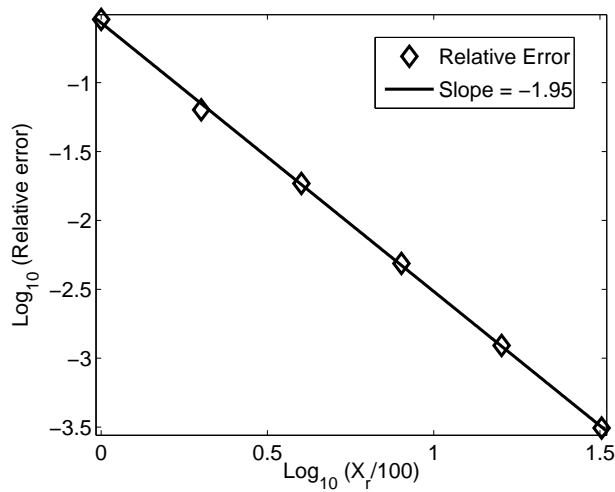}
  \caption{Total relative error $E_T$ of the energy density $U$ as a
    function of the grid size for an electric pulse crossing a
    dielectric interface. Here $X_r$ is the number of grid points, and
    the error is computed by $E_T$$=$$\frac{1}{L}\sum_{i=1}^{L} E_r$,
    with $L$ the grid size and $E_r$ the error at each cell location
    given by Eq.~\eqref{richardson2}.}\label{relativeerror}
\end{figure}

The theoretical predictions for the amplitudes of the transmitted and
reflected pulses can be computed from \cite{n16}
\begin{subequations}\label{OEMMD}
  \begin{equation}
    \frac{E_0'}{E_0}=\frac{2}{\sqrt{\frac{\epsilon_r'}{\epsilon_r}}+1} \quad ,
  \end{equation}
  \begin{equation}
    \frac{E_0''}{E_0}=\frac {\sqrt{\frac{\epsilon_r'} {\epsilon_r}}-1}
    {\sqrt{\frac{\epsilon_r'}{\epsilon_r}}+1} \quad .
  \end{equation}
\end{subequations}
They give us $\frac{E_0'}{E_0}$$=$$0.7751$ and
$\frac{E_0''}{E_0}$$=$$0.2249$.  The values from the simulation are
$0.7750$ and $0.2249$, respectively, that is the errors are smaller
than $1\%$. This simulation takes less than 30 ms in a standard PC.

Because we are dealing with conservative equations in the
differential form, we need to calculate the order of convergence of
the model. Let us track how does it change the accuracy of a physical variable when the resolution of the spatial grid $\delta x$ increases. The time resolution increases in the same way, because $\delta x = \sqrt{2} c \delta t$, with $c$ the speed of light. We choose to track the electromagnetic energy density $U$, defined by
\begin{equation}\label{eq:energydensity}
  U = \frac{1}{2}\left ( \epsilon_r \vec{E}\cdot \vec{E} + \frac{1}{\mu_r} \vec{B} \cdot \vec{B}  \right ) \quad .
\end{equation}
Fig.~\ref{energydensity} shows the electromagnetic
energy density $U$ as a function of the $z$-coordinate for different
grid sizes ($200$, $800$ and $3200$ grid points). 
We use Richardson's method \cite{rich1,rich2} to
compute the convergence error for the model. First, let us estimate the exact
solution of $U$ up to order $n$ by using the expression
\begin{equation}\label{richardson1}
  U = \lim_{\delta x \rightarrow 0} U(\delta x) \approx \frac{2^n U\left (\frac{\delta x}{2}\right ) - U(\delta x)}{2^n - 1} + \vartheta(\delta x^{n+1})\quad ,
\end{equation}
with errors $\vartheta(\delta x^{n+1})$ of order $n+1$. Here, we set $n=2$. Thus, the
relative error at each cell between the value $U(\delta x)$ and the ``exact''
solution $U$ is computed by
\begin{equation}\label{richardson2}
  E_r(\delta x) = \left |\frac{U(\delta x) - U}{U} \right | \quad .
\end{equation}
The total relative error $E_T$ is computed just by adding the errors $E_r$ on all cells. Fig.~\ref{relativeerror} shows that this error decreases as $\delta x^{1.95}$, supporting that the present scheme has a second-order convergence, i.~e. the convergence behavior we expected for our LB model.

\subsection{Skin Effect} \label{SectionSkin}

The skin effect is the exponential decay in the amplitude of a plane
wave penetrating a conducting medium. To reproduce the effect we
construct an uni-dimensional space of $L$ cells as before, with zero
conductivity for $z$$<$$L/4$ and $\sigma_0$ conductivity for
$z$$>$$L/4$. We choose an smooth conductivity transition of the form
\begin{equation}
  \sigma=\sigma_0(1 + \tanh(x-L/4) ) \quad ,
\end{equation}
to avoid numerical instabilities. The incoming plane wave is generated
by imposing a harmonic oscillation of the electric field at $z$$=$$0$,
\begin{subequations}\label{ondaplana}
\begin{equation}
  \vec{E}=(E_0 \sin(\omega t),0,0) \quad ,
\end{equation}
\begin{equation}
  \vec{B}=(0,B_0 \sin(\omega t),0) \quad ,
\end{equation}
\end{subequations}
where $E_0$ is the wave amplitude, $B_0$$=$$E_0/c$ and $\omega$ is the
angular frequency.

\begin{figure}
  \centering
  \includegraphics[angle=0,scale=0.32]{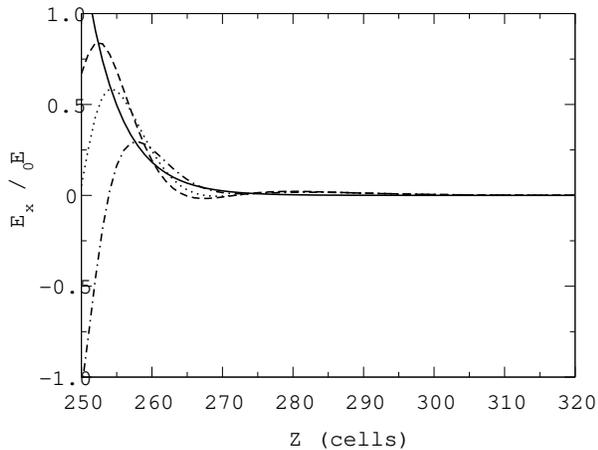}
  \caption{Numerical reproduction of the skin effect. A plane wave
    impacts perpendicular a conducting medium at $z$$=$$250$.  The
    dashed, dot and dot-dashed lines draw the electric field inside
    the conductor as a function of depth after $635$, $647$ and $670$
    time steps, respectively.  The solid line is the theoretical
    exponential decrease one expects for the amplitude of the incident
    wave. One can observe that the first maxima of the incident waves
    lie on the expected exponential curve.}\label{skin1}
\end{figure}

The simulation is shown in Figure \ref{skin1}, with $L$$=$$1000$,
$E_0$$=$$0.001$, $\omega$$=$$\pi/100$, $c$$=$$1/\sqrt{2}$ and
$\sigma_0$$=$$10^6$ (in normalized units).  The theoretical expression
for the {\bf amplitude} of the oscillating electric field inside the
conductor is
\begin{equation}\label{amortigua}
  E=E_0 \exp(-x/\Delta) \quad ,
\end{equation}
where $E_0$ is the amplitude of the electric field just outside and
$\Delta$ is known as the skin thickness. For good conductors this
thickness is given by \cite{n16}
\begin{equation}\label{efectoskinb}
  \Delta=\sqrt{\frac{2}{\sigma \mu \omega}} \quad .
\end{equation}
Figure \ref{skin1} also shows the analytical solution given by Eq.
(\ref{amortigua}) as a solid line. One can observe that the amplitude
of the electric field oscillation follows in excellent agreement the
theoretical prediction.  This simulation took less than 30 ms in a
single Pentium IV at 3.0 GHz.

\subsection{Electric Dipole}

\begin{figure}
  \centering
  \includegraphics[scale=0.5]{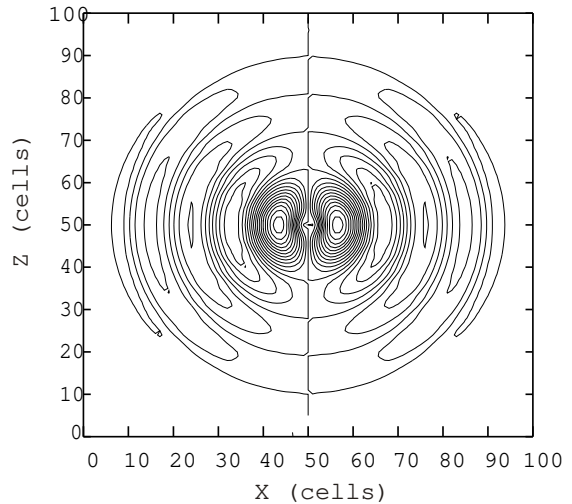}
  \caption{Lines of equal intensity of the magnetic field produced by an
    oscillating electric dipole in $z$ after two whole oscillations
    (approximately).  }\label{radiationsim}
\end{figure}
In order to simulate the radiation pattern of a small electric dipole
antenna we construct an array of $L\times L \times L$, with $L$$=$$100$
cells with free boundary conditions (each limit cell takes itself as
his own missing neighbor). In the center of this array we insert a
small oscillating current density in the $z$ direction,
\begin{equation}\label{dipcurrent}
  J_z = J'_0\sin \biggr( \frac{2\pi}{T} t\biggl ) \quad ,
\end{equation}
where $J'_0$ is the amplitude of the current density and $T$ is the
oscillation period. For avoiding any abrupt change of the physical
quantities between two neighboring cells we choose actually a Gaussian
functional form for the amplitude of the current density $J'_0$,
\begin{equation}\label{dipcurrent0}
  J'_0 = J_0 \exp(-0.75[(x-50)^2 +(y-50)^2+(z-50)^2]) \quad .
\end{equation}
The period was set to $T$$=$$25.0$ time steps and the amplitude to
$J_0$$=$$0.0001$ (in automaton units).

\begin{figure}
  \centering
  \includegraphics[scale=0.31]{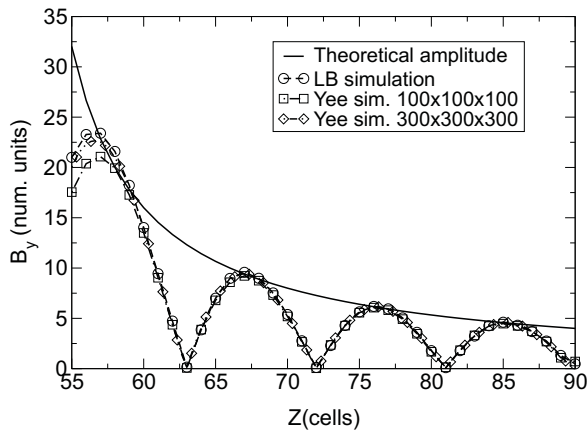}
  \caption{Amplitude along the $x$ axis of the oscillating magnetic
    field produced by the electric dipole of figure
    \ref{radiationsim}. The theoretical envelope corresponds to
    Eq.\eqref{radiacion}. The simulation results are obtained by using
    our LB model with an array of $100\times100\times100$ cells
    (circles) and by the method by Yee with resolutions of
    $100\times100\times100$(squares) and $300\times300\times300$
    (diamonds) grid points.}
  \label{dipolocomp}
\end{figure}

The results are shown in Figures \ref{radiationsim} and
\ref{dipolocomp}. Figure \ref{radiationsim} draws the lines of equal
intensity of the magnetic field after $56$ time steps. Figure
\ref{dipolocomp} shows the amplitude of the radiated magnetic field
along the $x$ axis, compared both with a set of results from Yee
simulations for several grid resolutions and with the theoretical
values for the peaks, given by \cite{n16}
\begin{equation}\label{radiacion}
  B=\frac{k^2 P}{x}\sqrt{1-\frac{1}{k^2x^2}} \quad.
\end{equation}
Here, $k$$=$$\frac{\omega}{c}$ is the magnitude of the wave vector,
with $c$ the speed of light in vacuum and $\omega$ the angular
frequency. The magnitude $P$ of the electric dipole momentum is
computed by
\begin{equation}\label{momentumd}
  P=\frac{J_0 \Sigma}{\omega} \quad ,
\end{equation}
where $\Sigma$ is the effective volume for the dipole. The Yee method
was implemented for three resolutions: $100\times 100\times 100$,
$200\times 200\times 200$, and $300\times 300\times 300$ grid points,
in contrast with the array of $100\times 100\times 100$ cells for our
LB method. Table \ref{tabledipole} shows the amplitude of the peaks
computed from the LB and Yee methods and contrasted the theoretical
values.

\begin{table}
  \centering
  \begin{tabular}{|c|c|c|c|c|}\hline
    Amp. LB &  Amp. Yee  & Amp. Yee & Amp. Yee & Theo. value\\
    $L = 100 $   &  $L=100$   &  $L=200$  &  $L=300$ &  $B_{theo}$\\
    $B_{sim, LB}$ & $B_{sim, Yee}$ & $B_{sim, Yee}$ & $B_{sim, Yee}$  &   \\ \hline
    23.33 & 21.00 & 22.18  & 22.62  & 23.01 \\ \hline
    9.53 &  9.21 & 9.38  & 9.38 & 9.37 \\ \hline
    6.12 &  6.05 & 6.09  & 6.15 & 6.07 \\ \hline
    4.54 &  4.47 & 4.56  & 4.54 & 4.51 \\ \hline
  \end{tabular}
  \caption{Amplitude of the peaks of Fig.~\ref{dipolocomp} computed form our LB model ($B_{sim, LB}$) with a resolution of $L=100$ cells and from the Yee's method ($B_{sim, Yee}$) with several resolutions. The theoretical values ($B_{theo}$) predicted by Eq.~(\ref{radiacion}) area also included.}
  \label{tabledipole}  
\end{table}

\begin{table}
  \centering
  \begin{tabular}{|c|c|c|c|c|}\hline
    Err. LB & Err. Yee & Err. Yee & Err. Yee  \\
    ($\%$) & ($L=100$), ($\%$) & ($L=200$), ($\%$) & ($L=300$), ($\%$)  \\ \hline
    1.4 & 8.7 & 3.6 & 1.6 \\ \hline
    1.7 & 1.7 & 0.1 & 0.1 \\ \hline
    0.8 & 0.3 & 0.3 & 1.3 \\ \hline
    0.7 & 0.9 & 1.1 & 0.7 \\ \hline
  \end{tabular}
  \caption{Relative errors on the amplitude of the peaks of Table \ref{tabledipole} as contrasted with the theoretical predictions.}
  \label{tabledipole2}  
\end{table}

The simulation using the LB model matches again the
theoretical predictions with an accuracy of less than $2\%$ (see
Table~\ref{tabledipole2}). In contrast, the Yee's method with the same grid
resolution computes the nearest peak to the source (where the boundary effects are less pronounced) with an error of $8.7\%$, far from the LB one. In order to reach the same accuracy that the LB method, we must increase the grid resolution by the Yee's method up to $L=300$, but at expenses of high computational costs. The simulation using the LB model takes $23$ seconds in a single standard machine, while the Yee's method takes $5$ seconds with $L=100$, $67$ seconds for $L=200$ and $336$ seconds for $L=300$, that is around $13$ times more CPU time that the LB for the same errors.

Finally, the results obtained in this section illustrate the
possibilities of our method to study the radiation by antennas.

\subsection{Microstrip Waveguide}
\begin{figure}
  \centering
  \includegraphics[scale=0.6]{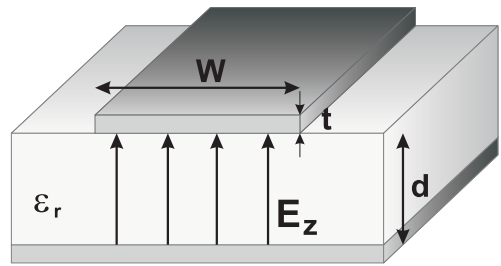}
  \caption{Microstrip waveguide. Here $W$ is the width of the upper
    metallic plate and $t$ its thickness, $d$ is the dielectric
    thickness between the two parallel plates with relative
    permittivity $\epsilon_r$ and $E_z$ is the electric input
    signal.}\label{guia}
\end{figure}

Let us simulate the wave propagation on a microstrip waveguide (see
figure \ref{guia}). For this purpose, we chose a 3D array of
$100$$\times$$50$$\times$$50$ cells with free boundary conditions and
we insert two parallel metallic layers of conductivity $\sigma$, one
wider than the other (figure \ref{guia}), in vacuum.

\begin{figure}
  \centering
  \includegraphics[scale=0.26]{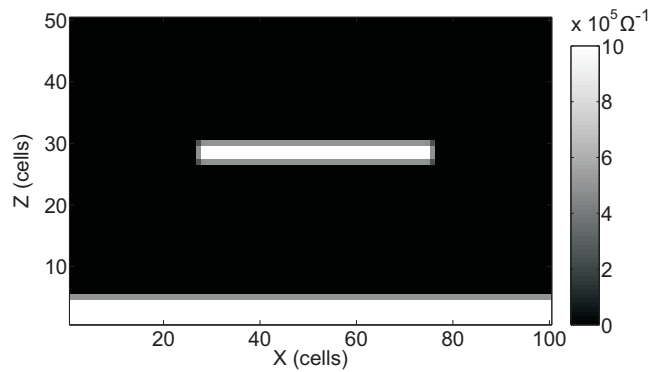}
  \caption{Electric conductivity as a function of $x$ and $z$ at
    $y$$=$$25$.}\label{conguia}
\end{figure}

The dimensions, depicted in figure \ref{guia}, are (in normalized
units): $t$$=$$5$, $d$$=$$20$ and $W$$=$$50$. The signal enters into
the waveguide by forcing the cells at $y$$=$$0$ to have the electric
and magnetic fields of a plane wave (Eqs. \eqref{ondaplana}) with
$E_0$$=$$0.01$, $c$$=$$1/\sqrt{2}$ and $\omega$$=$$\pi/100$ (see
figure \ref{guia}). The conductivity is smoothly changed along three
cells, as in Sec. \ref{SectionSkin} (see figure \ref{conguia}).

\begin{figure}
  \centering
  \includegraphics[scale=0.26]{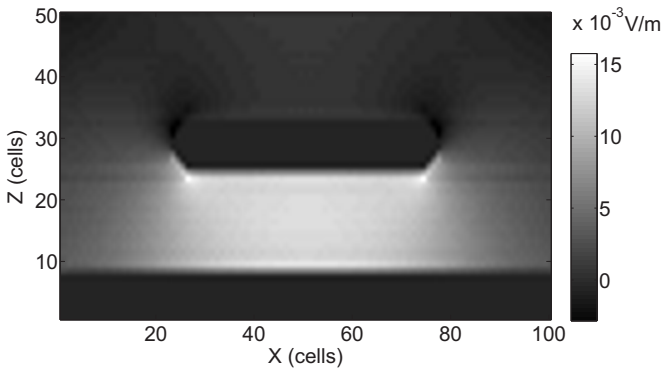}
  \caption{The $z$ component of the electric field in the microstrip
    at $y$$=$$25$.}\label{electricguia}
\end{figure}

\begin{figure}
  \centering \includegraphics[scale=0.26]{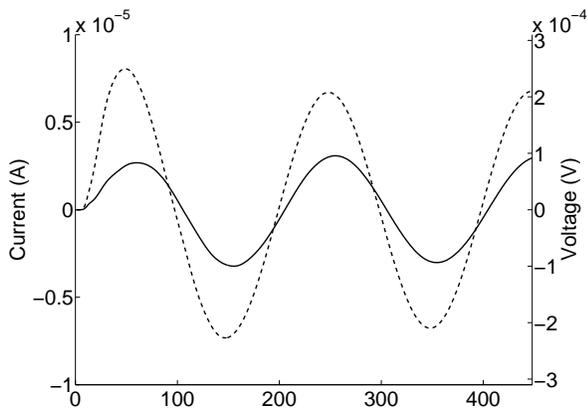}
  \caption{Voltage (dashed line) and current (solid line) in the microstrip
    waveguide at $y$$=$$25$ as a function of time.}\label{tensiont}
\end{figure}

Figure \ref{electricguia} shows the $z$ component of the electric
field at $y$$=$$25$ after $467$ time steps. The voltage $V(z,t)$ and
the current $I(z,t)$ along the waveguide at that time are drawn in
figure \ref{tensiont}. They are in phase, as expected for a pure
traveling wave.  The waveguide impedance, computed as $Z_0$$=$$V/I$,
gives us $Z_0$$=$$70.73$ $\Omega$. In contrast, the theoretical value
for the impedance of an infinite microstrip waveguide is given by
\cite{n32}
\begin{widetext}
  \begin{equation}\label{impedancia}
    Z=\frac{Z_0}{2\pi\sqrt{2(1+\epsilon_r)}}\ln{\biggl
      [1+\frac{4d}{\omega_{eff}}\biggl(\frac{14+\frac{8}{\epsilon_r}}{11}\frac{4d}{\omega_{eff}}+\sqrt{\biggl(
        \frac{14+\frac{8}{\epsilon_r}}{11}\frac{4d}{\omega_{eff}}
        \biggr)^2+\pi^2\frac{1+\frac{1}{\epsilon_r}}{2} } \biggr) \biggr] }
    \quad ,
  \end{equation}
  with
  \begin{equation}\label{anchoeff}
    \omega_{eff}=W+t\frac{1+\frac{1}{\epsilon_r}}{2\pi}\ln \Biggl [ 4e\Biggl(
    \biggl(\frac{t}{d} \biggr)^2+\biggl
    (\frac{1}{\pi}\frac{1}{\frac{W}{t}+\frac{11}{10}} \biggr)^2
    \Biggr)^{-\frac{1}{2}} \Biggr] \quad .
  \end{equation}  
\end{widetext}
From these equations we obtain a theoretical value of
$Z_0$$=$$72.6$$\Omega$. This gives us a $3$$\%$ difference between the
simulation and the theoretical prediction. The simulation takes 83
seconds in a single standard PC.

\subsection{Resonant Cavity}
As a last benchmark, we simulate a cubic resonant cavity and find its
resonance frequencies. The cavity is an array of
$50$$\times$$50$$\times$$50$ cells, corresponding to a size of
$5$$\times$$5$$\times$$5$ $cm$ with periodic boundary conditions, but
imposing a null electric field at the boundary (a perfect conductor).
An emitter point is set inside the cavity as a single cell with an
oscillating electric field in the $x$ direction (the antenna) and a
receptor point (the detector) is chosen as a single cell where we
measure the electric field amplitude.  The emitter point was set at
$(5,5,5)$ (in cell units) and the receptor point at $(5,5,45)$. The
oscillation frequency was scanned from $0.0182$ to $0.027$
oscillations per time step by multiplying each previous value by a
constant factor of $0.0072$, for a total of $22$ different
frequencies. Each frequency starts a single run, consisting of $5000$
time steps before equilibrium and four whole oscillations to estimate
the amplitude of the electric field.
\begin{figure}
  \centering
  \includegraphics[scale=0.34]{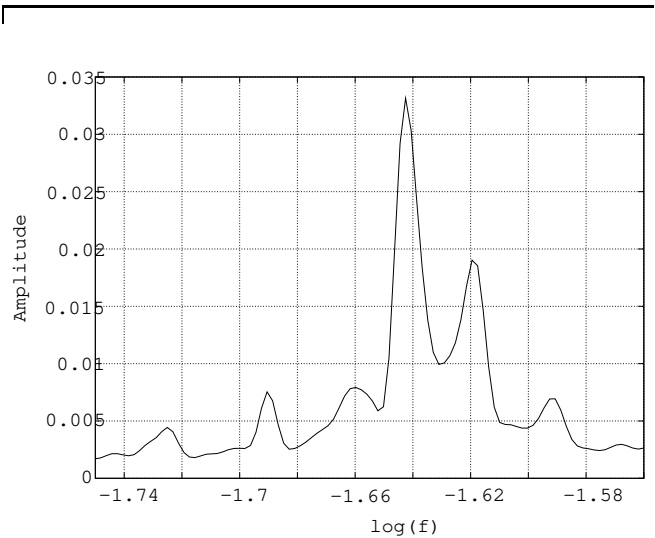}
  \caption{Frequency spectrum for the resonant cavity with size
    $50$$\times$$50$$\times$$50$ cells in the frequency range between
    $0.0182$$clicks^{-1}$ and $0.027$$clicks^{-1}$.}\label{espectro}
\end{figure}

\begin{table}
  \centering
  \begin{tabular}{|c|c|c|}\hline
    Experimental values & Theoretical values & Error \\ 
    $\log(f_{exp})$ & $\log(f_{the})$ & ($\%$) \\ \hline
    -1.725 & -1.724 &  0.06 \\ \hline
    -1.690 & -1.694 &  0.24 \\ \hline
    -1.660 & -1.669 &  0.54 \\ \hline
    -1.643 & -1.646 &  0.18 \\ \hline
    -1.618 & -1.625 &  0.43 \\ \hline
    -1.590 & -1.589 &  0.57 \\ \hline
  \end{tabular}\caption{Simulational $f_{exp}$ and theoretical $f_{the}$
    resonant frequencies for a  cavity of size $50$$\times$$50$$\times$$50$
    cells. The frequencies are in oscillations per time
    step.}\label{tablepicos}
\end{table}

Figure \ref{espectro} shows the amplitude of the electric field as a
function of frequency. The resonant peaks are clearly identified.
Table \ref{tablepicos} compares the values for the resonant
frequencies with the theoretical predictions computed by \cite{n16}
\begin{equation}\label{resonancia}
  \omega_{nmp}=\frac{c}{\epsilon_r
    \mu_r}\sqrt{\biggl(\frac{n\pi}{L_x}\biggr)^2+\biggl(\frac{m\pi}{L_y}\biggr)^2+\biggl(\frac{p\pi}{L_z}\biggr)^2
  } \quad .
\end{equation}
All differences are smaller than $1\%$. The whole simulation ($22$
runs of more than $5000$ time steps each on a space of $125000$ cells) takes $50$
minutes in a single standard PC.
  
\section{Discussions and Conclusions}
\label{conclusion}
In this manuscript, we introduce a three-dimensional LBGK model that
reproduces the Maxwell equations in materials.  The model successfully
accounts for the behavior of electromagnetic fields inside
dielectrics, magnets and conductors, and give us information about the
current density, electric charge and electromagnetic fields
everywhere.  It allows us to simulate a broad range of complex
phenomena with errors between $1$\% and $3$\% in all cases we tried,
namely: the amplitude of the transmitted and reflected pulses at a
dielectric interface, the exponential decay in the amplitude of an
incident plane wave inside a conductor (that is, the skin effect), the
amplitude (as a function of distance) of the magnetic field radiated
by an electric dipole, the characteristic impedance of a microstrip
waveguide and the resonant frequencies of a cubic resonant cavity.
These five benchmarks prove that our LB model works pretty well in the
more different situations.

The model uses D3Q13 velocity vectors, but assigns four auxiliary
vectors and four density functions to each one of them. This gives us
a total of $33$ vectors and $50$ density functions per cell. These
auxiliary vectors allow for the explicit construction of any
conservation law with curls, like the Faraday's and Amp\`ere's laws
\cite{PRL}; a procedure that has also been successful for the
construction of a LBGK reproducing the two-fluids model of
magnetohydrodynamics \cite{nmiller}, but in our case, the density
functions dealing with the electromagnetic fields ought to be doubled
to account for the Maxwell equations in media.  This increase in
complexity allows for the more diverse applications, even with
realistic values of dielectric, magnetic and conductivity constants.

The equilibrium distribution functions,
\begin{equation}
  f_{ij}^{p(r) \rm eq}(\vec{x},t)=\frac{1}{16}\vec{v}_i^p \cdot
  \vec{J'}+\frac{\epsilon^{1-r}}{4}\vec{E'} \cdot
  \vec{e}_{ij}^{p}+\frac{\mu^{r-1}}{8}\vec{B} \cdot \vec{b}_{ij}^{p} \quad ,
\end{equation}
with $r$$=$$0,1$, deserve a special discussion. What does these
equilibrium function actually mean? If we consider
$\vec{e}_{ij}^{p}$ and $\vec{b}_{ij}^{p}$ as small fluctuations in the
electromagnetic fields, $\vec{E}$$=$$\vec{E}+\vec{e}_{ij}^{p}$ and
$\vec{B}$$=$$\vec{B}+\vec{b}_{ij}^{p}$, the increment in the energy
density, at first order, is
\begin{equation}
  \Delta U(\vec{E},\vec{B})=\epsilon (\vec{E}\cdot\vec{e}_{ij}^{p}) +
  \frac{1}{\mu}( \vec{B}\cdot\vec{b}_{ij}^{p}) \quad,
\end{equation}
which resembles our expression for the equilibrium distribution
functions $f_{ij}^{p(0) \rm eq}$. So, we argue that such equilibrium
functions can be interpreted as changes in the electromagnetic energy
density propagating along the velocity vectors $\vec{v}_i^p$.
Moreover, because the auxiliary vectors $\vec{e}_{ij}^{p}$ are related
to the electric field and the vectors $\vec{b}_{ij}^{p}$ related to
the magnetic field, the velocity vectors
$\vec{v}_i^p$$=$$\vec{e}_{ij}^{p} \times \vec{b}_{ij}^{p}$ can be
linked to the direction of the Poynting vector, in some sense.

The model allows us to select the electromagnetic constants (magnetic
permeability, electric permittivity and conductivity) for each cell at
pleasure. It just needs to smooth the transition between two different
materials across three cells, approximately, in order to avoid
numerical instabilities (this is the limit of stability), but this is
a standard procedure in many numerical models \cite{n100, n37}.  For
these reasons, our LB seems to be promising in a variety of
applications, including among others: the propagation of
electromagnetic pulses inside microcircuits produced by atmospheric
rays or discharge antennas, the electromagnetic diffraction across
objects with complex geometries or the electromagnetic propagation
across meta-materials.  They can also be an excellent alternative for
the design and optimization of antennas, which requires to simulate a
large number of configurations, and very specially in the design of
pulse antennas, where single-frequency numerical methods fail. All
these applications can be theme of future works.

In terms of speed, our model is one order of magnitude faster than the
original Yee's FDTD method\cite{n100, yee2, yee3} to obtain the same
accuracy level on the simulation of the electromagnetic fields
produced by an oscillating electrical dipole. Actually, the grid
resolution of the Yee's method must be increased three times in order
to reach the same accuracy than our LB method. It is interesting to
discuss why it could be. Yee's method uses six variables per cell
(three for the electric field and three for the magnetic one), but
each cell has also to look at the variables of all six first neighbors
to evolve in time, gathering a total of $42$ floating point variables
to evolve each cell at every time step. For comparison, our LB method
uses $50$ floating point variables per cell (the distribution
functions), i.~e. a similar number of variables to process, but it
takes around five times longer to evolve a LB cell than a FDTD cell by
the method of Yee.  On terms of memory consuming, our LB uses $50$
floating point variables per cell, that is eight times more memory per
cell than Yee's method, but this last requires three times more
spatial resolution to reach the same accuracy than our LB; therefore,
Yee's method needs three times more memory ($6$ variables per cell
times $27$ cells) to reach the same accuracy than our LB.  It seems,
therefore, that our LB pack the information on the electromagnetic
fields in a more condensed way than FDTD, although the floating point
operations at every time step are more complex. In addition, each cell
in our LB encloses all the information it needs to evolve into the
{\it same} cell, before it pass the results to the neighbors; in
contrast, each cell at FDTD needs to borrow the information from its
neighbors {\it before} it can compute its new values. This difference
can be exploited for the efficient implementation of our LB on
multigrid computers, where LB methods in general have shown a
great performance\cite{parallel1, parallel2}. Of course, we are
comparing our LB method with the most basic form of Yee's FDTD, but
the same is true for our LB, that is still in its most basic form.
Many improvements has been done on FDTD along the years, many of them
related with non-uniform grid sizes or similar geometrical
improvements. We expect that similar improvements can be done on our
LB model in the future.

Our LB model for electrodynamics works fine using a relaxation time
$\tau$$=$$\frac{1}{2}$. However, in analogy with LB for fluids, this
regime must be unstable when the system is very far from the
equilibrium. In the case of LB models for fluids, there are extensions
using the H-theorem to improve the stability and allows to study e.~g. turbulent 
systems\cite{ELB1}. Because our model follows the same Boltzmann equation and has all
the same characteristics, it is valuable to think on the possibility
of making an Entropic LB for electrodynamics that would improve the
results for very complex systems far from equilibrium. This will be a subject of future developments. Another interesting issue is the modeling of non-linear current or charge-density terms, that can be easily included in the sources. How the LB responds to this change would be a nice area of future work.

Hereby we have introduced a Lattice-Boltzmann model for
electrodynamics that actually reproduces the Maxwell equations in
media, with a plenty of future applications. The model shows itself an
order of magnitude more efficient than the original FDTD method by
Yee, and employs realistic values of the electromagnetic constants
describing the media. Moreover, it also illustrates how to construct
three-dimensional LB models that fulfills conservation laws with
antisymmetric tensors. We hope that this valuable theoretical
development will push forward the evolution of LB models further away
in the horizon of even more exciting applications.

\begin{acknowledgments}
  The authors are thankful to Dominique d'Humi\`eres, Sauro Succi and
  Paul Dellar for very fruitful discussions and to Hans J. Herrmann
  for hospitality and help.  We thank the graduate scholarship program
  of the National University of Colombia and the Colombian Excellence
  Center for the Simulation and Modeling of Complex Systems,
  CeiBA-Complejidad for financial and travel support. We are also deeply thankful to
  two anonymous referees for they valuable suggestions and
  corrections, which have strongly enhanced this manuscript.
\end{acknowledgments}

\appendix
\section{Chapman-Enskog Expansion} 
\label{ChapmanEnskog}

The Boltzmann equations (Eq. \eqref{lbe2} and \eqref{lbe4}) determine
the system evolution. This evolution rule gives in the continuum limit
the macroscopic differential equations the system satisfies.  In
order to determine such macroscopic equations we develop a
Chapman-Enskog expansion, as follows. Let us start by taking the Taylor
expansion of the Boltzmann equations until second order in spatial and
temporal variables,
\begin{eqnarray}{\label{lbee1ED}}
  \begin{aligned}
    &\vec{v}_i^p \cdot \vec{\nabla} f_{ij}^{p(r)} +\frac{1}{2}
    \sum_{\alpha,\beta} \frac{\partial^2 f_{ij}^{p(r)} }{\partial x_\alpha
      \partial x_\beta}(v_{i\alpha}^p v_{i\beta}^p) \\& +\frac{\partial
      f_{ij}^{p(r)}}{\partial t} + \frac{\partial}{\partial t}\vec{v}_i^p
    \cdot \vec{\nabla} f_{ij}^{p(r)} \\ &+ \frac{1}{2}\frac{\partial^2
      f_{ij}^{p(r)}}{\partial t^2}\delta t^2 =
    -\frac{1}{\tau}(f_{ij}^{p(r)}-f_{ij}^{p(r)\rm eq}) \quad ,
  \end{aligned}
\end{eqnarray}
\begin{eqnarray}{\label{lbee3ED}}
  \begin{aligned}
  \frac{\partial f_{0}^{(r)}}{\partial t}+\frac{1}{2}\frac{\partial^2
    f_{0}^{(r)}}{\partial t^2}=-\frac{1}{\tau}(f_{0}^{(r)}-f_{0}^{(r)\rm eq})
  \quad ,
  \end{aligned}
\end{eqnarray}
where $\alpha, \beta$$=$$x, y, z$ denote the $x$, $y$ and $z$
components.

Next, we expand the distribution functions and both the spatial and temporal
derivatives into a power series of a small parameter, $\lambda$,
\begin{equation}
  f_{ij}^{p(r)}=f_{ij}^{p(r)(0)}+\lambda f_{ij}^{p(r)(1)}+\lambda^2
  f_{ij}^{p(r)(2)}+... \quad ,
\end{equation}
\begin{equation}
  \frac{\partial}{\partial t}=\lambda \frac{\partial}{\partial t_1}+\lambda^2
  \frac{\partial}{\partial t_2}+... \quad ,
\end{equation}
\begin{equation}
    \frac{\partial}{\partial x_\alpha}=\lambda \frac{\partial}{\partial
      x_{\alpha 1}}+... \quad .
\end{equation}
It is assumed that only the $0$th order terms of the distribution
functions contribute to the macroscopic variables. So, for $n>0$ we
have
\begin{subequations}{\label{nomacrosED}}
  \begin{equation}
    \sum_{i,j,p} f_{ij}^{p(r)(n)} \vec{e}_{ij}^p=0 \quad ,
  \end{equation}
  \begin{equation}
    \sum_{i,j,p} f_{ij}^{p(r)(n)} \vec{b}_{ij}^p=0 \quad .
  \end{equation}
\end{subequations}

The main current density $\vec{J'}$ is of the order $\lambda$ \cite{n30}, so we can write $\vec{J'}$$=$$\lambda \vec{J'}_1$. Because $f_{ij}^{p(r)\rm eq}$ is now a function of $\vec{Jq'}$, we need to develop a Chapman-Enskog expansion of the equilibrium function,
\begin{equation}
  f_{ij}^{p(r)\rm eq}=f_{ij}^{p(r)(0)\rm eq}+\lambda f_{ij}^{p(r)(1)\rm eq}+\lambda^2 f_{ij}^{p(r)(2)\rm eq} \quad .
\end{equation}

If we replace these results into Eqs.\eqref{lbee1ED} and \eqref{lbee3ED}, we
obtain for the zeroth order in $\lambda$
\begin{subequations}{\label{zeroth}}
\begin{equation}
  f_{ij}^{p(r)(0)\rm eq}=f_{ij}^{p(r)(0)} \quad ,
\end{equation}
\begin{equation}
  f_{0}^{(r)\rm eq}=f_{0}^{(r)(0)} \quad .
\end{equation}
\end{subequations}
For the first order in $\lambda$ we gather
\begin{subequations}{\label{firstED}}
\begin{eqnarray}{\label{firstbED}}
  \begin{aligned}
    \vec{v}_i^p \cdot \vec{\nabla}_1 f_{ij}^{p(r)(0)} &+\frac{\partial f_{ij}^{p(r)(0)}}{\partial t_1} = \\& 
    -\frac{1}{\tau}(f_{ij}^{p(r)(1)}-f_{ij}^{p(r)(1)\rm eq}) \quad ,  
  \end{aligned}
\end{eqnarray}
\begin{eqnarray}{\label{firstdED}}
  \begin{aligned}
 \frac{\partial f_{0}^{(r)(0)}}{\partial
 t_1}=-\frac{1}{\tau}(f_{0}^{(r)(1)}-f_{0}^{(r)(1)\rm eq}) \quad ,  
  \end{aligned}
\end{eqnarray}
\end{subequations}
and for the second order in $\lambda$ we have
\begin{subequations}{\label{secondED}}
\begin{eqnarray}{\label{secondbED}}
  \begin{aligned}
  &\biggl(1-\frac{1}{2\tau}\biggr)\biggl(\vec{v}_i^p \cdot \vec{\nabla}_1+\frac{\partial }{\partial
  t_1}\biggr)f_{ij}^{p(r)(1)} \\&+ \frac{\partial f_{ij}^{p(r)(0)}}{\partial
  t_2} + \frac{1}{2\tau}\biggl(\vec{v}_i^p \cdot \vec{\nabla}_1+\frac{\partial }{\partial
  t_1}\biggr)f_{ij}^{p(r)(1)\rm eq}= \\&-\frac{1}{\tau}(f_{ij}^{p(r)(2)}-f_{ij}^{p(r)(2)\rm eq}) \quad ,
  \end{aligned}
\end{eqnarray}
\begin{eqnarray}{\label{seconddED}}
 \frac{\partial f_{0}^{(r)(0)}}{\partial t_2}=
 -\frac{1}{\tau}(f_{0}^{(r)(2)}-f_{0}^{(r)(2)\rm eq})   \quad . 
\end{eqnarray}
\end{subequations}

The first order and the second order terms for the equilibrium
function of the electromagnetic fields are obtained by replacing the
Eq. \eqref{expandE} into Eq.\eqref{equilc}. We obtain
(gathering together the same powers of $\lambda$) ,
\begin{subequations}{\label{equilcesED}}
\begin{eqnarray}
  f_{ij}^{p(r)(0)\rm eq}(\vec{x},t)=\frac{\epsilon^{1-r}}{4}\vec{E} \cdot
  e_{ij}^{p}+\frac{\mu^{r-1}}{8}\vec{B} \cdot b_{ij}^{p} \quad ,
\end{eqnarray}
\begin{eqnarray}
  f_{ij}^{p(r)(1)\rm eq}(\vec{x},t)=\frac{\lambda}{16}\vec{v}_i^p \cdot
  \vec{J'}_1-\frac{\lambda \mu_0}{16}\vec{J'}_1 \cdot e_{ij}^{p} \quad ,
\end{eqnarray}
\begin{eqnarray}
  f_{ij}^{p(r)(2)\rm eq}(\vec{x},t)=0 \quad ,
\end{eqnarray}
\begin{eqnarray}
  f_{0}^{(r)(0)\rm eq}(\vec{x},t)=\rho_c \quad ,
\end{eqnarray}
\begin{eqnarray}
  f_{0}^{(r)(1)\rm eq}(\vec{x},t)= f_{0}^{(r)(2)\rm eq}(\vec{x},t)=0 \quad .
\end{eqnarray}
\end{subequations}

Now, we are ready to determine the equations the model satisfies
in the continuum limit. First, let us consider $\tau$$=$$1/2$. By
adding Eqs. \eqref{firstbED}, \eqref{firstdED}, \eqref{secondbED}
and \eqref{seconddED} over $i$, $j$ and $p$, we get
\begin{eqnarray}{\label{Max00ED}}
  \frac{\partial \rho_c}{\partial t_1}=0 \quad ,
\end{eqnarray}
and
\begin{eqnarray}{\label{Max01ED}}
  \frac{\partial \rho_c}{\partial t_2}+\nabla \cdot \vec{J'}_1 =0 \quad .
\end{eqnarray}
Summing up these two equations gives
\begin{eqnarray}{\label{Max02ED}}
  \frac{\partial \rho_c}{\partial t}+\nabla \cdot \vec{J'} =0 \quad .
\end{eqnarray}

Multiplying the equations \eqref{firstbED}, \eqref{firstdED},
\eqref{secondbED} and \eqref{seconddED} by $\vec{e}_{ij}^p$ and
summing up over the index $i$, $j$ and $p$ gets for $r$$=$$0$
\begin{eqnarray}{\label{Max10ED}}
  \frac{\partial (\epsilon \vec{E})}{\partial t_1}-\frac{1}{2}\vec{\nabla} \times
  \biggl(\frac{\vec{B}}{\mu}\biggr)=-\frac{1}{2} \mu_0 \vec{J'}_1 \quad ,
\end{eqnarray}
and
\begin{eqnarray}{\label{Max11ED}}
  \frac{\partial (\epsilon \vec{E})}{\partial t_2}-\frac{\mu_0}{4}\frac{\partial
  \vec{J'}_1}{\partial t_1}=0 \quad . 
\end{eqnarray}
If we add these two equations, taking into account Eq.
\eqref{expandE}, we arrive to the first Maxwell equation,
\begin{eqnarray}{\label{Max12ED}}
  \frac{\partial (\epsilon \vec{E'})}{\partial t}-\frac{1}{2}\vec{\nabla}
  \times \biggl(\frac{\vec{B}}{\mu}\biggr)=-\mu_0 \frac{1}{2} \vec{J'} \quad .
\end{eqnarray}
Similarly, multiplying the Eqs. \eqref{firstbED} and \eqref{secondbED}
by $\vec{b}_{ij}^p$ and summing up over $i$, $j$ and $p$ for 
for $r$$=$$1$ gives
\begin{eqnarray}{\label{Max20ED}}
  \frac{\partial \vec{B}}{\partial t_1}+\vec{\nabla} \times \vec{E}=0 \quad ,
\end{eqnarray}
and
\begin{eqnarray}{\label{Max21ED}}
  \frac{\partial \vec{B}}{\partial t_2}-\frac{1}{2}\vec{\nabla} \times
  \biggl(\frac{1}{2} \mu_0 \vec{J'}_1 \biggr)=0 \quad .
\end{eqnarray}
Adding these two equations gives us the second Maxwell equation,
\begin{eqnarray}{\label{Max22ED}}
  \frac{\partial \vec{B}}{\partial t}+\vec{\nabla} \times \vec{E'}=0 \quad .
\end{eqnarray}

The others two Maxwell equations can be obtained from the
Eqs.\eqref{Max12ED} and \eqref{Max22ED} as follows\cite{n3}: 
Let us apply the divergence operator to both sides of Eqs. \eqref{Max21ED} and
\eqref{Max22ED} to obtain
\begin{eqnarray}{\label{Max30ED}}
  \frac{\partial (\vec{\nabla} \cdot \vec{E'})}{\partial t}=-\frac{1}{2} \mu_0
  \vec{\nabla} \cdot \vec{J'} \quad ,
\end{eqnarray}
\begin{eqnarray}{\label{Max40ED}}
  \frac{\partial (\vec{\nabla} \cdot \vec{B})}{\partial t}=0 \quad .
\end{eqnarray}
Because of the continuity equation, Eq.\eqref{Max02ED}, 
\begin{eqnarray}{\label{Max32ED}}
  \begin{aligned}
    \frac{\partial (\vec{\nabla} \cdot \vec{E'})}{\partial t}=\frac{1}{2}
    \mu_0 \frac{\partial \rho_c}{\partial t} \quad .
  \end{aligned}
\end{eqnarray}
Thus, we obtain
\begin{eqnarray}{\label{Max33ED}}
  \begin{aligned}
    \frac{\partial (\vec{\nabla} \cdot \vec{E'}-\frac{1}{2} \mu_0 \rho_c )}{\partial t}=0 \quad .
  \end{aligned}
\end{eqnarray}
Summarizing, if the initial conditions of the electromagnetic fields satisfies the
Maxwell equations
\begin{eqnarray}{\label{Max41ED}}
  \vec{\nabla} \cdot \vec{B}=0 \quad ,
\end{eqnarray}
\begin{eqnarray}{\label{Max34ED}}
  \begin{aligned}
    \vec{\nabla} \cdot \vec{E'}=\frac{1}{2} \mu_0 \rho_c =
    \frac{\rho_c}{\epsilon_0} \quad ,
  \end{aligned}
\end{eqnarray}
these equations will be reproduced for all times. This
way to include the Gauss law and the null divergence of the magnetic field is well known in the literature.
It has been reported \cite{sonnendrucker} that employing this procedure to reproduce both the Maxwell equations and the motion of charged particles in a self-consistent way adds numerical errors in the discrete form of the charge-conservation equation. 
But this is not our case, because we are not solving the motion equation of any charged particle;
we are just solving the Maxwell equations with sources.

Finally, equations \eqref{Max12ED}, \eqref{Max22ED},
\eqref{Max41ED} and \eqref{Max34ED} determine the evolution of the electromagnetic
fields. These are the electrodynamics equations for
macroscopic media that the lattice Boltzmann model reproduces in the
continuum limit, with second order accuracy in space and time. This
completes the proof.
 
\bibliography{tesisV2}

\end{document}